\documentclass[twocolumn,showpacs,preprintnumbers,prb,aps]{revtex4-1}
\usepackage{hyperref}
\pdfoutput=1

\usepackage[pdftex]{graphicx}
\usepackage{dcolumn}
\usepackage{amsfonts,amsmath,amssymb,bm}
\usepackage{etoolbox} 
\usepackage{subfigure}

\begin{document}

\title{\bf {Graphyne as the anode material of magnesium-ion batteries: ab initio study }}
\date{\today} 
\author{E.~Shomali$^a$}
\author{I.~Abdolhosseini~Sarsari*$^{a,b}$}
\author{F.~Tabatabaei$^a$}
\author{MR.~Mosaferi$^a$}
\author{N.~Seriani$^c$}
\affiliation{a)~Department of Physics, Isfahan University of Technology, Isfahan, 84156-83111, Iran\\
b)~Computational Physical Sciences Research Laboratory, School of Nano-Science, Institute for Research in Fundamental Sciences (IPM), P.O. Box 19395-5531, Tehran, Iran\\
c) The Abdus Salam ICTP, Strada Costiera 11, 34151 Trieste, Italy\\
}

\begin{abstract}

Graphyne, a single atomic layer structure of the carbon six-member rings connected by one acetilenic 
linkage, is a promising anode of rechargeable batteries. In this paper, a first-principle study has 
been carried out on graphyne as a new candidate for the anode material of magnesium-ion batteries, 
using density functional theory calculations. The main focus is on the magnesium adsorption on graphyne 
surface. The structural properties such as adsorption height and energy, the most stable adsorption 
sites, the Band structure and DOS of the pristine graphyne the diverse Mg-decorated graphyne structures, 
and energy barrier against Mg diffusion are also calculated. As a consequence of the band structure and 
DOS of graphyne structures, it is found that the pristine graphyne and the Mg-decorated graphyne 
structures show a semiconducting nature and metallic behavior, respectively. Moreover, the migration 
behavior of Mg on graphyne for the main diffusion paths is determined by Nudged Elastic Band (NEB) 
method.

\end{abstract}

\maketitle
\section{INTRODUCTION}

Nowadays, with the acceleration of global energy demands, the depletion of fossil fuels, and their 
environmental impacts; interest in renewable energies has increased considerably over the last 
decades. In the meantime, efforts to develop energy storage devices as a renewable energy 
technology become more prevalent. One of the most popular energy storage system, lithium-ion 
batteries, have received remarkable attention. A battery is an electrochemical device which 
converts chemical energy directly into electrical energy and consists of three components, i.e., 
the anode (negative electrode), the cathode (positive electrode), and electrolyte. During discharge, 
lithium ions depart from the anode to the cathode through the non-aqueus liquid electrolyte and 
carry the current which leads to passing electrons around the external circuit to power various 
systems. When the battery is charging, an external electrical power source make the current to 
transfer in the reverse direction and lithium ions migrate to 
the anode across the electrolyte \cite{XianxiaYuan,scrosati2010lithium}.
\\
Recently, rechargeable batteries have entered our daily life and have been widely applied in 
portable and telecommunication electronic devices, transportation, and household-type items. 
Basically, rechargeable batteries have a significant role in the development of electronic 
devices such as; hybrid electric vehicles, notebook, mobile, flash lights, radios, watches, etc. 
They became very common because of their high energy density, enhanced rate capabilities, low 
self-discharging, long cycle life, excellent safety features, and etc. A great deal of attention 
has been paid to Li-ion battery technologies which are conquering the market, but due to having a 
better chance of reaching maturity, we decide to apply a different material for this technology 
\cite{meng2012recent}\cite{etacheri2011challenges}.
\\
Since the batteries coming into existence, graphite-based materials have been proposed for 
anodes. Graphite, the most stable form of carbon, has a layered, planar structure. The carbon 
atoms in each layer, have been placed in a honeycomb lattice and the individual layers are called 
graphene. Both graphite and graphene are allotropic forms of carbon. It is well known that 
graphite has a low capacity and that is not good enough for providing endless needs of the 
current society. Graphene, a two dimensional (2D) sheet of sp$^2$-hybridized carbon, has a higher 
capacity than graphite, mainly because lithium can be adsorbed on both sides of graphene, thus 
lithium has a larger space to diffuse and it can be led to higher rate of diffusion 
\cite{wu2011first}\cite{liu2016origin}. 
Graphene has many extraordinary properties which have attracted the attention of all researchers 
for a while. It is very strong and conducts heat and electricity efficiently \cite{neto2009electronic}. 
However, nowadays scientists are investigating that 
\textquotedblleft Could graphyne be better than graphene?\textquotedblright Graphyne (Gy), a 
novel carbon allotrope, is the single atomic layer structure of the carbon six-member rings 
connected by one acetylenic linkage ($-C\equiv C-$), which introduces a diverse range of optical 
and electronic properties. Moreover, the thing that makes graphyne special is the presence of 
both sp and sp$^2$ hybridized carbon atoms. Also, it can be said that graphyne has two different 
Dirac cones lying slightly above and below the Fermi level which means it is self-doped and 
naturally includes conducting charge carriers (electrons and holes) without any external doping. 
Therefore, graphyne is a promising semiconductor which can be used in electronic devices 
\cite{sarkar2015first}\cite{peng2014new}.
\\
In this paper, we have focused on graphyne and Mg-decorated graphyne-based systems. Magnesium, an 
inexpensive and environmentally friendly metal with great raw material abundance and good operational 
safety, is a relatively new option for use as an anode 
material in rechargeable batteries \cite{yang2011first}. (To begin, we examined stability of graphyne by 
calculating the phonon spectrum using FHI-aims package). Then, we have conducted our study finding 
possible sites of Mg adsorption on graphyne by calculating the adsorption energy and its corresponding 
adsorption height. Moreover, we examined the electronic structure of graphyne and Mg-decorated graphyne
systems at different sites by calculating the band structure and DOS spectrum. Finally, the main target 
of this work; study of in-plane and out-plane diffusion paths of Mg on a graphyne layer and out-plane 
diffusion pathway of Mg in bulk graphyne, corresponding energy curves of adsorption sites of Mg on 
graphyne monolayers, and corresponding transition states (TS); was determined using Nudged Elastic Band 
(NEB) method.

\section{Computational Details}

Our electronic structure calculations and geometry optimizations have been performed in the 
framework of density functional theory (DFT) by using plane wave pseudo-potential 
as well as numerical orbital - full potential techniques,  implemented in the Quantum Espresso 
\cite{Giannozzi2009} and FHI-aims \cite{Blum2009} computational packages, respectively. We used 
GGA-PBE exchange-correlation functional \cite{PhysRevLett.77.3865}.
The 7$\times$7$\times$1 Monkhorest-pack scheme was used for sampling first Brillouin zone. A vacuum thickness 
of about 22 Bohr was utilized to avoid interaction of adjacent molecules. Energy cutoffs of 35 
Ry and 350 Ry, were used for plane wave expansion of wave functions and electron density. 
The Nudged Elastic Band (NEB) method \cite{Caspersen2005}, implemented in Quantum Espresso, was 
also used to determine the reaction path.
In this method, a path with the lowest energy is determined for the reaction then a string of 
images of the system are created. Finally, the images are connected together through the 
hypothetical springs with the same spring constant. It results in an energy for each image. 
In this research, minimum Energy Path was calculated as a function of adsorption sites of Mg on 
Gy monolayer by considering 59 and 19 images for in-plane and out-plane diffusion and 23 images for 
diffusion pathway of Mg in bulk graphyne, respectively. 
The vdW-DF, non-local correlation functional, proposed by Dion et al. \cite{Dion2004} utilized to calculate distances 
between layers and distance between ion and layer.

One of the most important issues in condensed matter physics and in theoretical chemistry is the 
determination of the reaction path. Many different methods have been presented for finding 
Minimum Energy Paths (MEPs) and saddle points like the Drag method, the NEB method, the CPR 
method, the Ridge method, the DHS method, and the Dimer method; which the Nudged Elastic Band 
(NEB) method was used in this study. In the NEB method, which initial and final states of the 
reaction are known, a string of images of the system are created and connected together through 
the hypothetical springs with the same spring constant from the reaction configuration, R, to the 
product configuration, P. In fact, images of the system are the intermediate configurations of 
system. Afterward, an optimized algorithm is applied to relax the images and put them in the 
MEP. By using NEB method, it is possible to determine Minimum Energy Path (MEP), saddle point 
energy (maximum potential energy along the MEP), activation energy barrier, estimated transition 
rate and also the configuration of atoms during the transition 
\cite{henkelman2002methods} \cite{jonsson1998nudged}.

\section{RESULTS AND DISCUSSIONS}

To examine graphyne as a reliable anode material of magnesium-ion battery, we not only look at the 
pristine graphyne and its properties, but also focus on the adsorption behavior of the magnesium atoms 
onto the graphyne surface. We first relaxed a single-layer graphyne, which is displayed in Fig.\ref{band-dos}. 
The bond length between the sp- and sp$^2$-hybridized C atoms are 1.22 (\AA{}) and 1.42 (\AA{}), 
respectively and the C\textendash C bond length that connect sp- and sp$^2$-hybridized C atoms is 1.39 
(\AA{}) which is in agreement with other researchers \cite{zhang2011high}.
\\

\begin{figure}[!ht]
\centering
\includegraphics*[width=1\columnwidth]{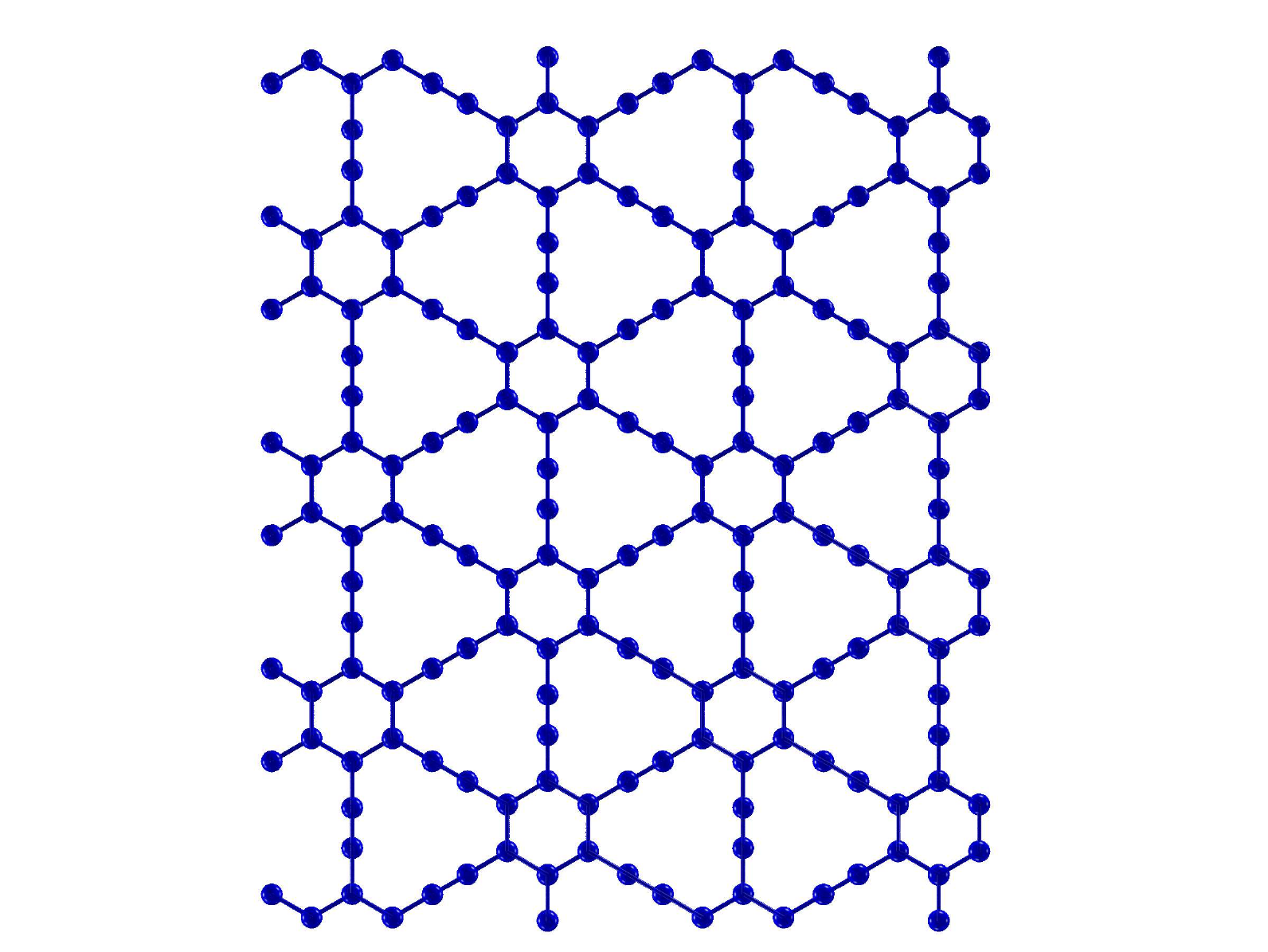}
\caption{ 
  Optimized structure of single-layer graphyne. The blue spheres represent C atoms.
}
\label{band-dos}
\end{figure}

\subsection{Mg adsorption on graphyne surface}

To survey the adsorption behavior, we considered the most stable locations for Mg decoration in graphyne. 
The Mg atom can adsorb above the centers of large and small hexagons which consisting of 12 and 6 carbon 
atoms, respectively and they are denoted as H- and h-sites hereafter. Another adsorption site for Mg 
named H-beside-site, H-bond-site, and h-bond-site in which Mg atom stands right above one of the 
corners of large hexagons, right above one of the carbon bonds of large hexagons, right above one of the 
carbon bonds of small hexagons, respectively. The different Mg-decorated graphyne structures are shown in 
Fig.~\ref{band-dos}.


We turn to investigate the electronic properties of graphyne with Mg atom in different positions. 
Table~\ref{Table1} summarizes the calculated binding energy, $(E_b)$, the distance between Mg and
graphyne, $(d_{Mg-graphyne})$, and the charge transfer of the Mg atom, where $E_b$ and $(d_{Mg-graphyne})$ 
are defined by equation (1) and van der Waals correction\cite{Dion2004}, respectively.
The $d_{Mg-graphyne}$ in h-bond-site, H-bond-site and h-site structures are 4.15(\AA{}), 4.02(\AA{}), and 4.07(\AA{}), respectively,
while in the case of the H-site structure, the distance between Mg and graphyne decreases to 1.34(\AA{}).
Among the four structure, the H-site
is the most favorable due to more negative binding energy in comparison to others.
The charge transfer from Mg atom to graphyne, is calculated by Bader analysis\cite{henkelman2006fast}. 
The results in \ref{Table1} show that as the Mg distance decreases the charge transfer increase. Mg atom in
H-site structure transfers 1.35 electrons to the substrate and More than others. 
Which would be responsible for the greater negative binding energy and consequently more stable structure.
\\
\begin{table}[h]
\centering
\caption{Calculated binding energies $(E_b)$, distances between Mg and graphyne $d_{Mg-graphyne}$ and the charge transfer of Mg atom}
\begin{tabular}{c  c   c  c}\hline\hline
\textbf{Structures}     &   $E_b (eV)$   &   distance(\AA{})   &   $\Delta Q|e|$    \\    [1ex]
\hline
\textbf{h-bond-site}   &     -0.47           &   4.15                    &   0.07                  \\    [0.5ex]
\hline
\textbf{H-bond-site}  &     -0.48           &    4.02                   &    0.08                  \\   [0.5ex]
\hline
\textbf{h-site}            &     -0.47           &   4.07                    &   0.07                  \\    [0.5ex]
\hline
\textbf{H-site}           &      -0.87          &   1.34                    &   1.35                  \\    [0.75ex]
\hline\hline
\label{Table1}
\end{tabular}
\end{table}
\\

\begin{align}
 E_b=(E_{Gy}+nE_{Mg}-E_{Mg-Gy})/n
\end{align}

To approach reality, a supercell containing two graphyne layers with an AB stacking sequence was also 
considered. The interlayer spacing in graphyne was set to be 3.45 (\AA{}), which has been recognized in 
the range of the most stable interlayer distances \cite{narita2000electronic}. The top view of bulk 
graphyne with an AB stacking sequence is shown in Fig.~\ref{NEB2}-a  . The Mg atom can be intercalated into the bulk 
graphyne in two sites, which are named HH- and hH-sites. In HH-site the Mg atom is take placed between 
centers of large hexagons of two adjacent layers and in hH-site the Mg atom is placed between centers 
of a small hexagon and a large hexagon of two adjacent layers. The binding energy of HH- and hH-sites are 
... and ..., respectively. The corresponding adsorption height for HH-site is about 1.8 (\AA{}), in which 
the Mg atom is nearly in the middle of the spacing of the two adjacent layers. In the hH-site, Mg atom is 
closer to the large hexagon with a distance of 1.24 (\AA{}). Comparison reveals that the intercalated Mg
prefers to occupy an H-site rather than other sites, due to its larger binding energy.
\\
It is worth noting that due to the existence of a van der Waals force between the Mg atom and the 
graphyne surface, the van der Waals effects have been considered in the above-mentioned calculations.
\\
\begin{figure*}[ht]
\centering
\includegraphics*[width=2\columnwidth]{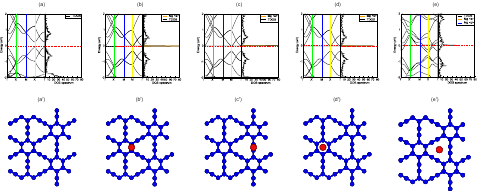}
\caption{The band structure of a) Gy, b) h-bond-site, c)H-bond-site, d) h-site and e) H-site. 
The Optimized structure of single-layer graphyne labled with prime. The blue and red spheres represent C atoms and Mg atom, respectively.}
\label{band-dos}
\end{figure*}

\subsection{Electronic properties}

The band structure and density of state (DOS) have been calculated to study the electronic properties of 
pristine graphyne and Mg adsorbed graphyne layer systems. The band structure, and Dos of the pristine 
graphyne and the different Mg-decorated graphyne structures are shown in 
Fig.~\ref{band-dos}. For pristine graphyne, the band structure and Dos imply a semiconducting 
behavior with the direct band gap of 0.358 (eV) due to the same location of the valence band maximum 
(VBM) and conduction band minimum (CBM) of graphyne. This gap is in good agreement with the results of 
0.38 (eV) for Na-decorated graphyne batteries from other DFT calculations. Moreover, it can be seen from 
the band structure and Dos of the Mg-decorated graphyne structures that there is a comparable downward 
shift of the conduction band for each structure (and, alternatively, an upward shift of the Fermi level), 
indicating that the Mg adsorption makes graphyne metallic. These results are consistent with the density 
of state of all structures considered in this paper. It is noteworthy that the downward shift of the 
conduction band for H-site is the highest among all structures considered in this research.

The calculated band structure and density of state of graphyne by replacing Mg atom on the top of the large 
and small hexagons (H and h sites) are presented in Fig.~\ref{band-dos} which exhibit an 
overlapping between conduction band and Fermi energy. All the other Mg-decorated graphyne structures show 
the same metallic nature but the amount of overlapping between the conduction band and Fermi energy is 
different \cite{sarkar2015first}.\\
In order to determine the constituents of the electronic band, the projected density of states was calculated for
graphyne with different Mg-decoration.  The state near the Fermi level are mainly dominated by Mg s orbital in h-bond-site, H-bond-site, and h-site.
Decreasing distance between Mg and substrate in H-site structure results in  the Mg s orbital contribution reduction in the
Fermi state. It is found that the Mg p and s orbitals have influenced in Fermi level, But the graphyne states play a very
important role in the Fermi state. The charge transfer on the substrate in this decoration is extremely more than others.

There is a large difference in electronegativity between Mg and C atoms which cause a charge transfer 
from Mg to C atoms and thus the metallic behavior can be seen, that is an important characteristic for 
some applications such as electrode materials in rechargeable batteries 
\cite{seyed2015theoretical} \cite{liu2015first}.

\subsection{Mg diffusion in Graphyne}

\begin{figure*}
   \centering
   \subfigure[][]{
      \label{h1-H1-H2-h2}
      \includegraphics[width=0.6\columnwidth]{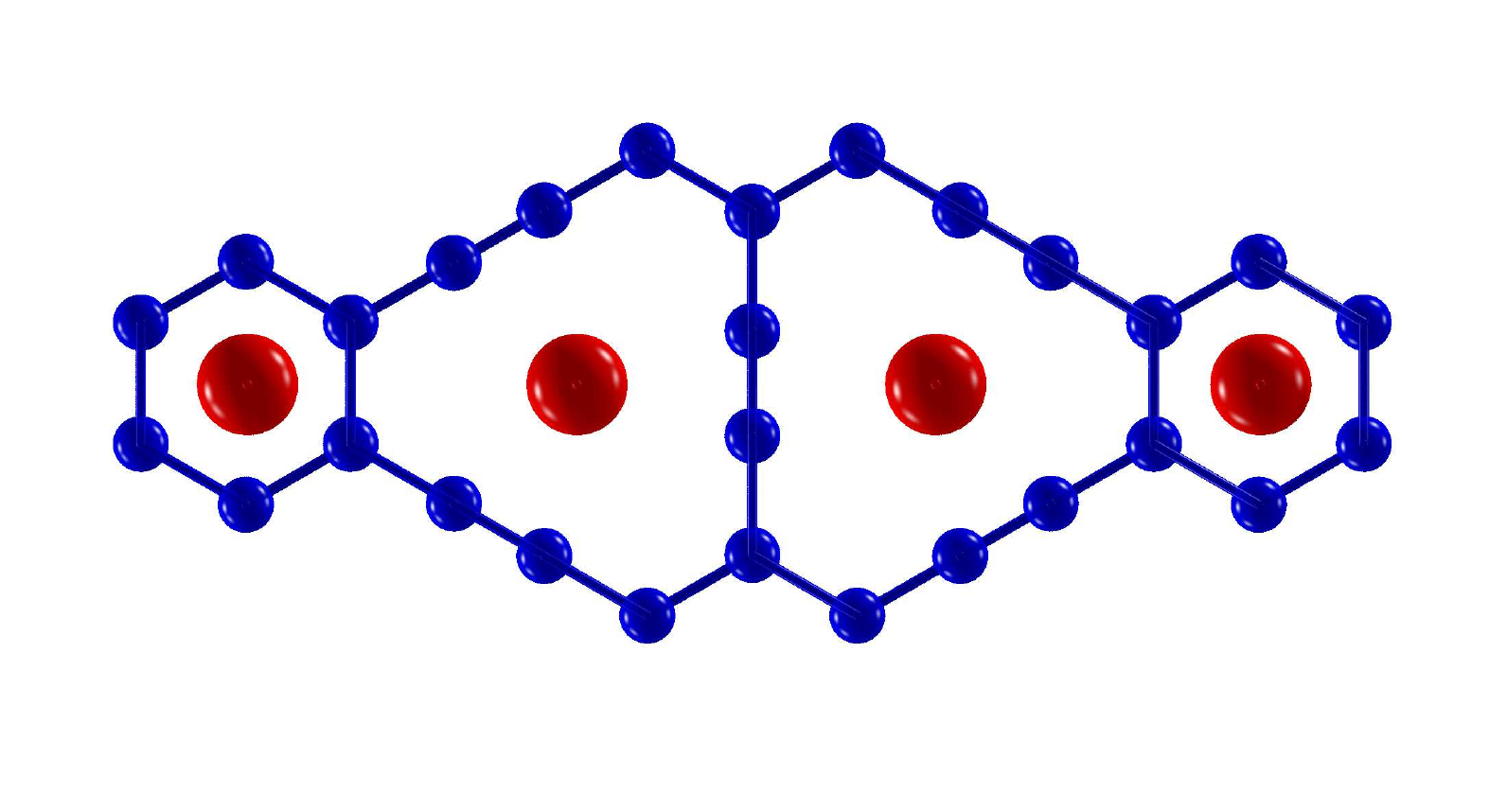}}
      \hspace{0.05\textwidth}
      \subfigure[][]{
      \label{h1-H1-h2-H2-up}
      \includegraphics[width=0.6\columnwidth]{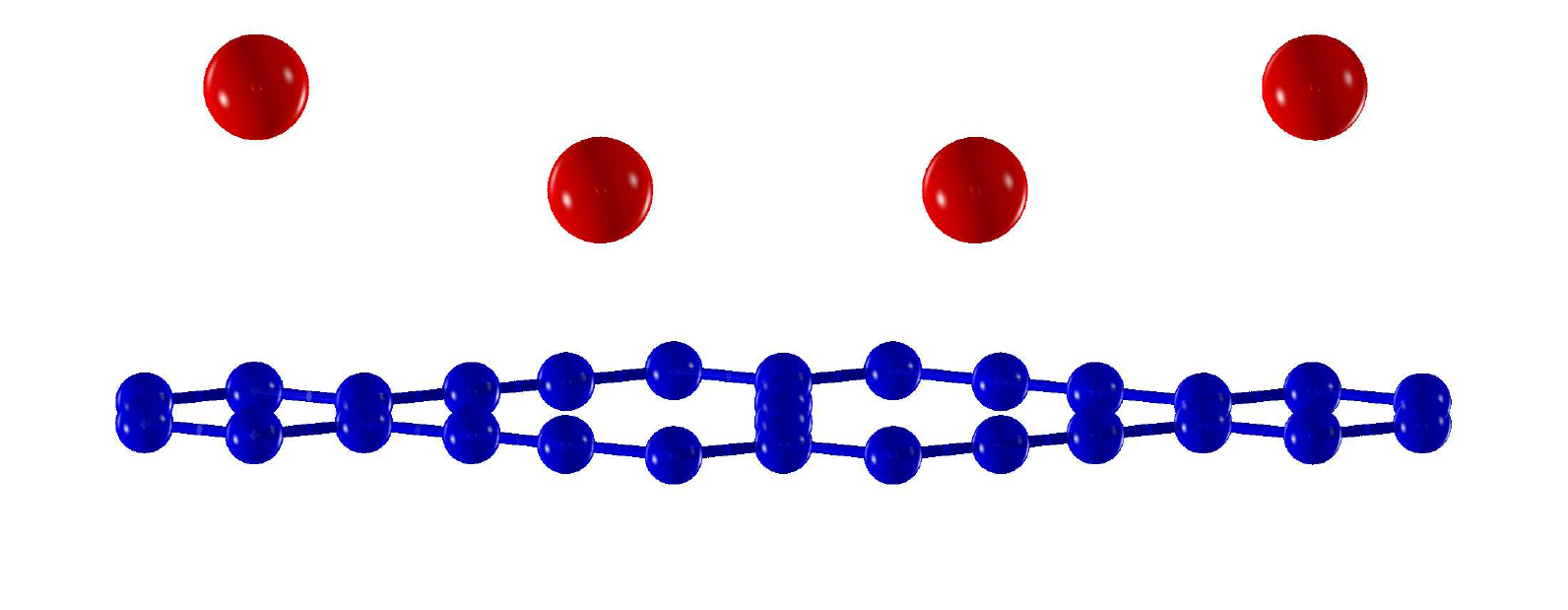}}
      \hspace{0.05\textwidth}
      \\
   \subfigure[][]{
      \label{H-up-down}
      \includegraphics[width=0.6\columnwidth]{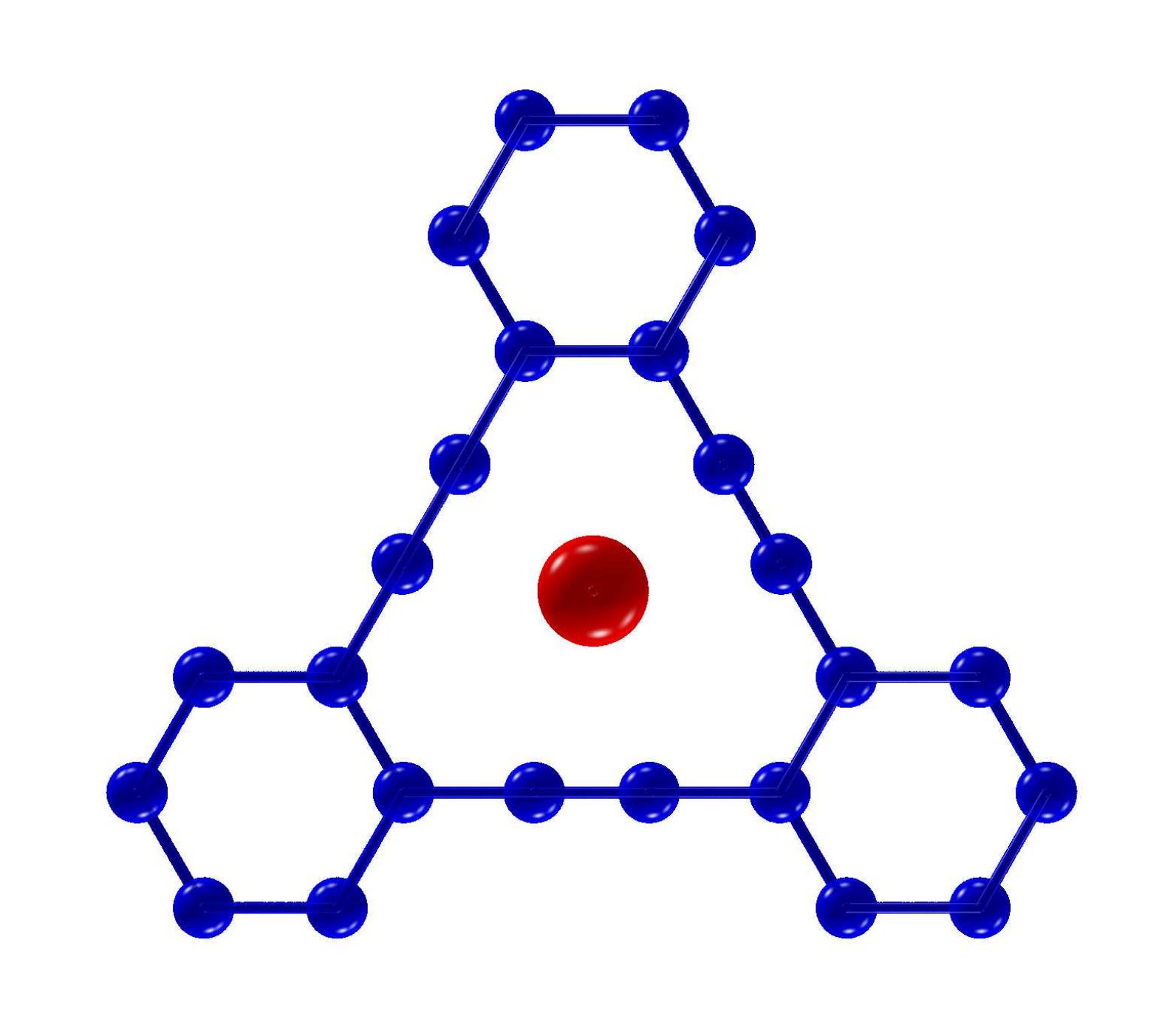}}
   \hspace{0.05\textwidth}
   \subfigure[][]{
      \label{h-up-down}
      \includegraphics[width=0.6\columnwidth]{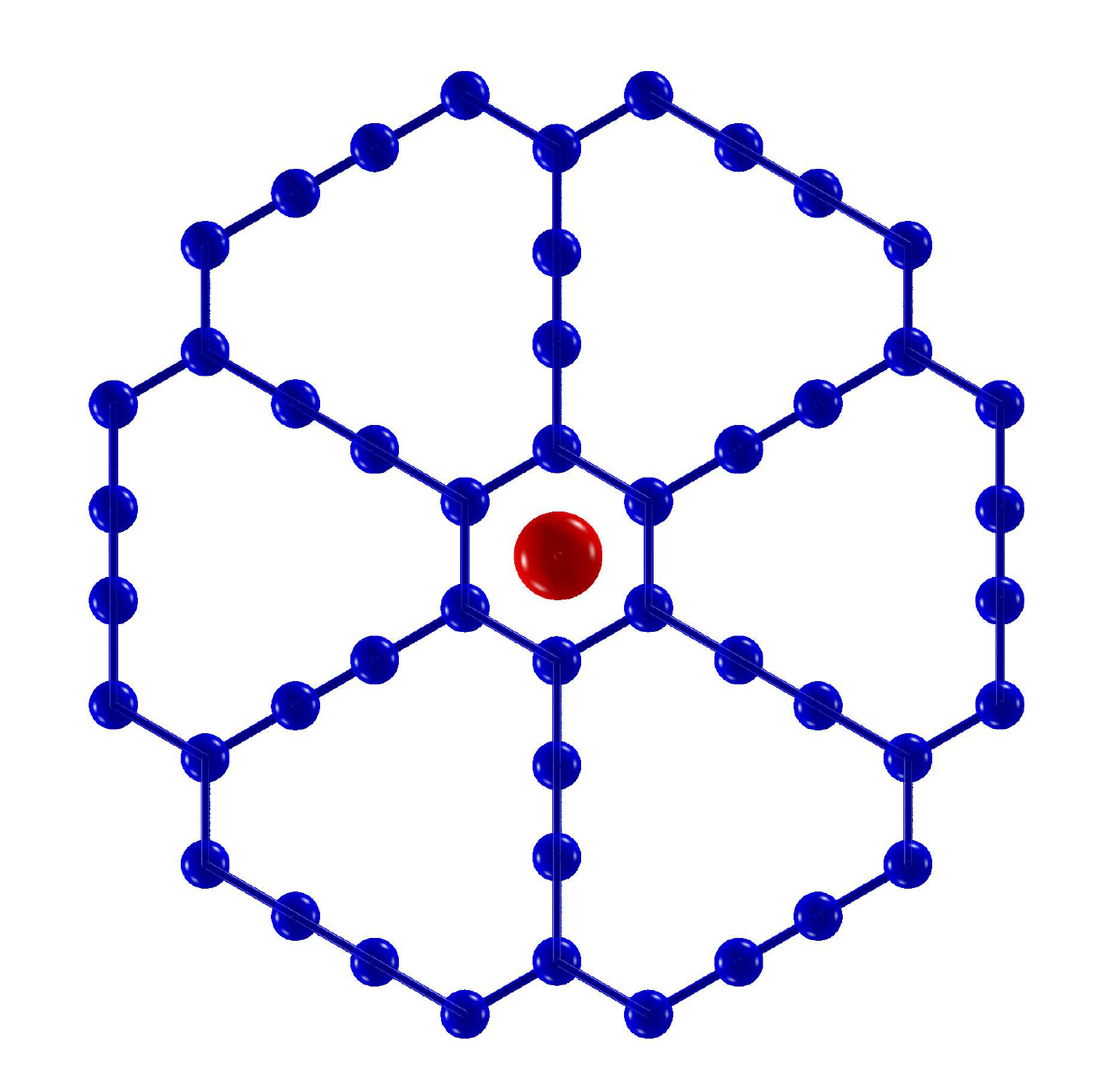}}
    \hspace{0.05\textwidth}
            \caption{ The schematic drawings of 
            a,b) in-plane diffusion paths (h1-H1-H2-h2) Side image and image from above 
            and two out-plane diffusion paths
            c) across the hollow of H-site
            d) across the hollow of h-site
}
\label{NEB}
\end{figure*}

\begin{figure*}
   \centering
   \subfigure[][]{
      \label{59-energy-h1-H1-H2}
      \includegraphics[width=0.8\columnwidth]{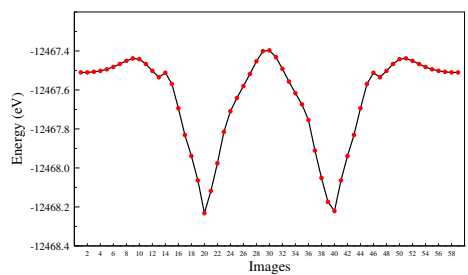}}
      \hspace{0.05\textwidth}
      \\
   \subfigure[][]{
      \label{19-H-up-down}
      \includegraphics[width=0.8\columnwidth]{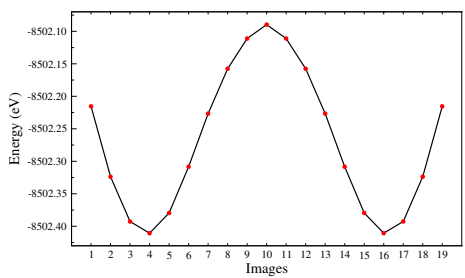}}
   \hspace{0.05\textwidth}
   \subfigure[][]{
     \label{19-h-up-down}
      \includegraphics[width=0.8\columnwidth]{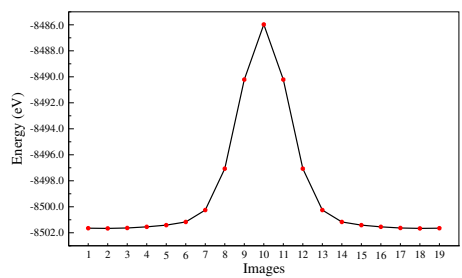}}
    \hspace{0.05\textwidth}
            \caption{ Mg migration behaviors
            a) in-plane diffusion paths (h1-H1-H2-h2)
            and two out-plane diffusion paths
            b) across the hollow of H-site
          c) across the hollow of h-site
}
\label{MEP}
\end{figure*}

\begin{figure*}
   \centering
   \subfigure[][]{
      \label{path-2_009}
      \includegraphics[width=0.7\columnwidth]{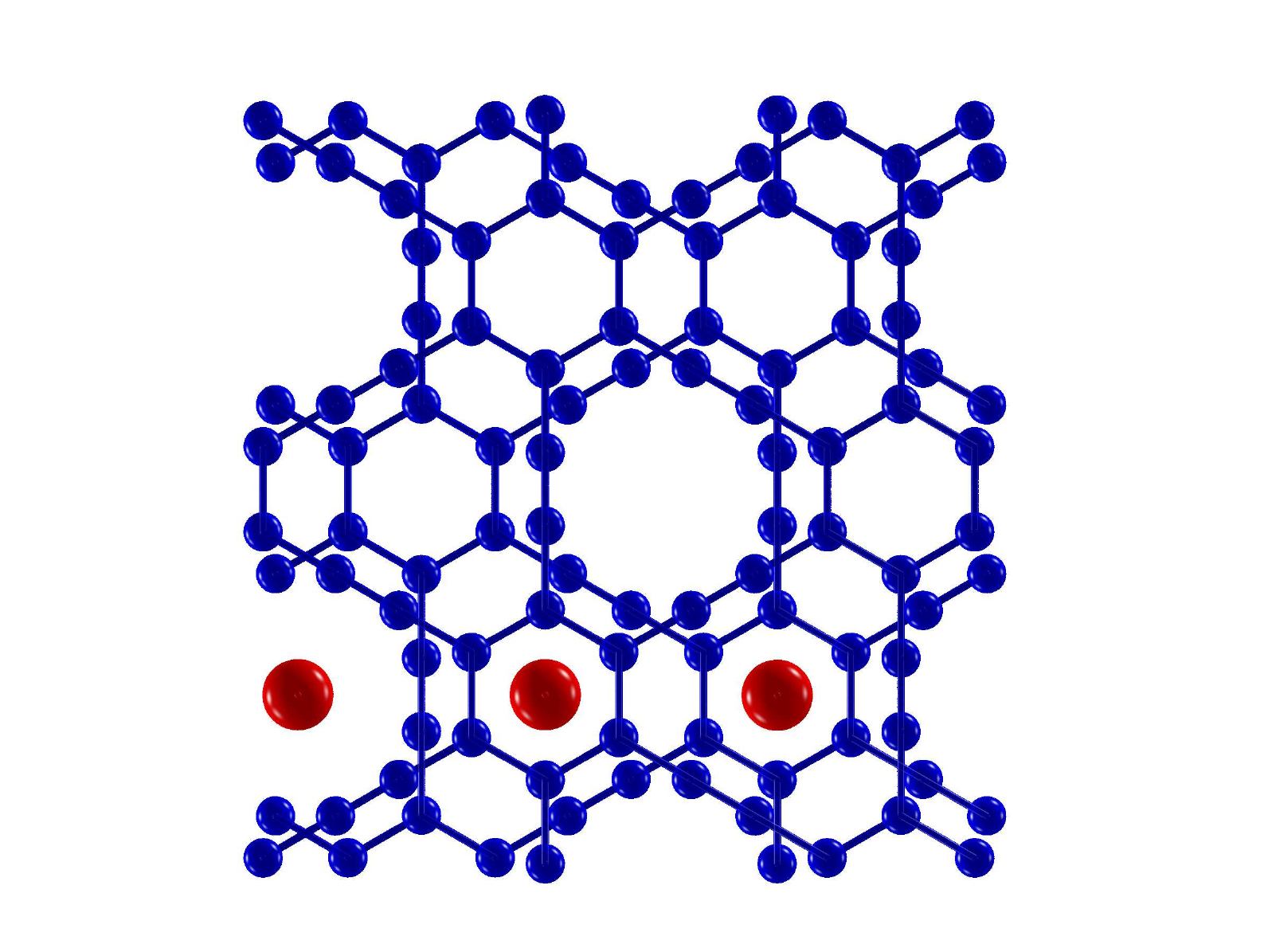}}
      \hspace{0.05\textwidth}
     \subfigure[][]{
      \label{path-2-up_015}
      \includegraphics[width=0.7\columnwidth]{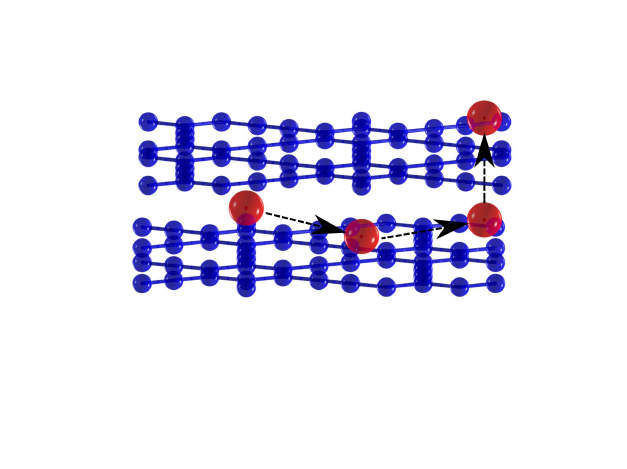}}
      \hspace{0.05\textwidth}
      \subfigure[][]{
      \label{energy-Hh-hH-HH-HH}
      \includegraphics[width=0.7\columnwidth]{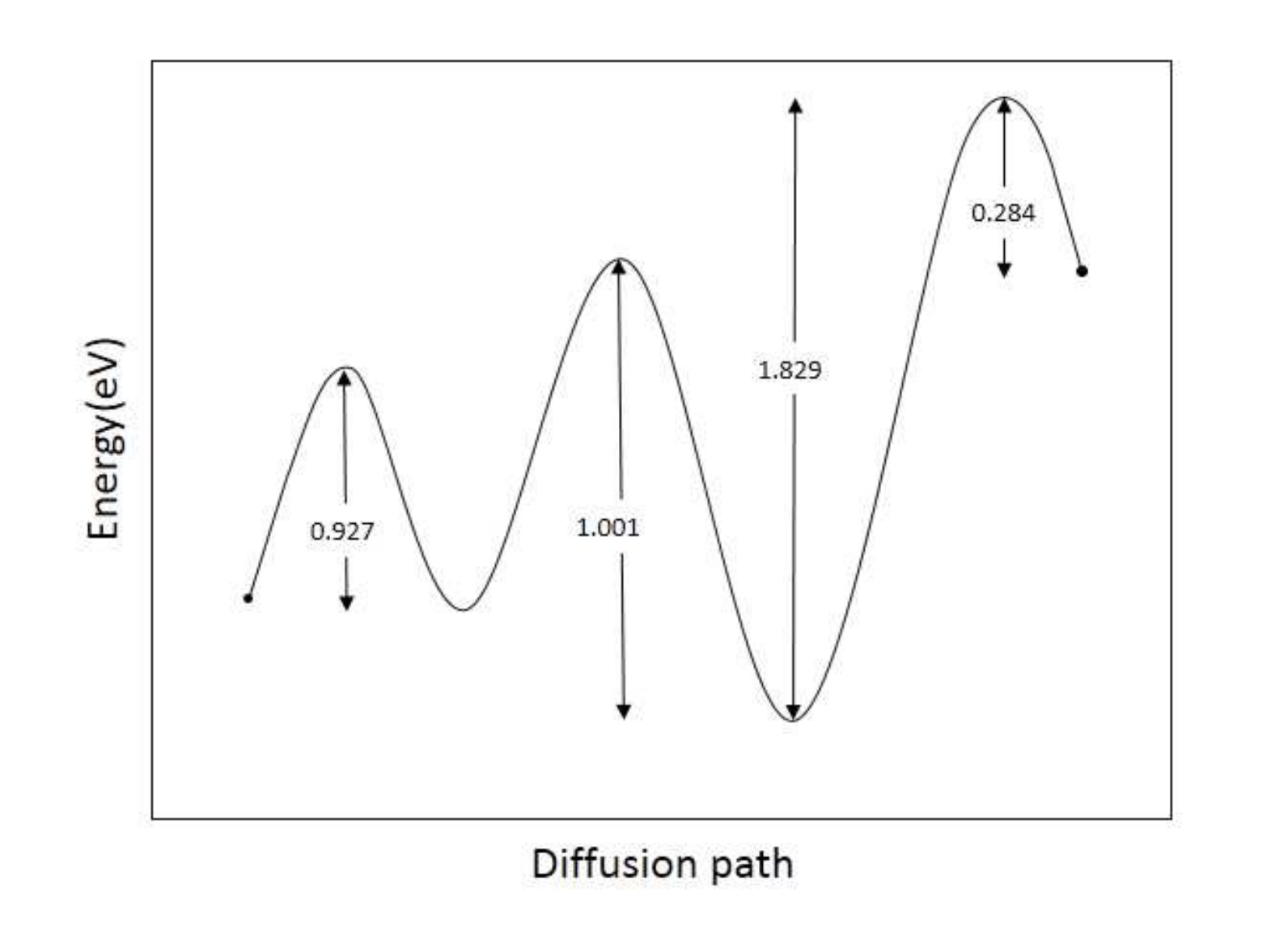}}
      \hspace{0.05\textwidth}
            \caption{ The schematic drawings of 
            a,b) out-plane diffusion pathway of Mg in bulk graphyne (Hh-hH-HH-HH) Side image and image 
            from above and b) the corresponding energy profile as a function of adsorption sites.
}
\label{NEB2}
\end{figure*}

Considering the significance of Mg mobility on graphyne monolayer, the diffusion path is calculated on 
the way to the most stable adsorption sites using the Nudged Elastic Band (NEB) method. We investigated 
Mg migration behaviors for two main diffusion paths; in-plane diffusion paths (h1-site to H1-site, 
H1-site to H2-site, H2-site to h2-site) and out-plane diffusion paths (across the hollow of H-site and 
h-site) as depicted in Fig.~\ref{NEB}, respectively. Since the calculation of energy curves is meaningful 
and very determinative on battery applications, the corresponding Minimum Energy Paths (MEPs) are also 
shown as a function of adsorption sites of Mg on graphyne monolayer and two layers in Fig.~\ref{MEP}. 
Moreover, the Mg migration behavior in a supercell containing two graphyne layers with an AB stacking 
sequence, for the main diffusion path (Hh-hH-HH1-HH2) was investigated which is shown in Fig.~\ref{NEB2}.

The results from Fig.~\ref{59-energy-h1-H1-H2} indicate that the activation energy barrier of Mg 
diffusion along the h-site to H-site direction is identified to be 0.072 (eV) and the transition state 
(TS) is located on the image.~9; where is near the top of the midpoint of the C(sp$^2$)-C(sp$^2$) bond between two 
adjoining hollows of large and small hexagons but more leaning to the large hexagon. Moreover, Mg 
migration between two neighboring H-site is as high as 0.835 (eV), more than that in h-site to H-site 
diffusion path, and it can be seen that the transition state (TS) is specified near the top of midpoint 
of the C(sp)$\equiv$C(sp) bond between two adjacent hollows of large hexagons; where is known by 
image.30. As the reaction pathway is followed, there is a high energy barrier of about 0.783 (eV) for Mg 
diffusion along the H-site to h-site path; which is reside on image.51 and represent the energetic 
favorability of the H-site over the h-site. In addition, the h1-site to H-site diffusion path with the 
lowest energy barrier of about 0.072 (eV) is mainly dominated in the in-plane Mg migration on Gy.
\\
In the following, the energy profiles of two typical out-plane diffusion paths are also studied, which 
can be seen in Fig.~\ref{19-H-up-down} and \ref{19-h-up-down}, respectively. In out-plane diffusion paths 
Mg migrates from one side of a graphyne layer to another side along the direction perpendicular to the 
graphyne plane. The energy barrier for Mg diffusion through a large and small hexagon is about 0.32 (eV) 
and 15.68 (eV), respectively. Therefore, Mg can easily migrate through the large hexagon by overcoming a 
relatively small energy barrier of 0.32 (eV). Moreover, it can be seen that the out-plane diffusion of 
Mg across the small hexagon is energetically and kinetically prohibited because of such a high energy 
barrier. Eventually, the large hexagon (H-site), formed by the sp$^2$- and sp-hybridized C atoms in 
graphyne, is recommended for both Mg adsorption and out-plane Mg migration. In addition, according to 
Fig.~\ref{energy-Hh-hH-HH-HH} the energy barrier of Mg diffusion along the Hh-site to hH-site and hH-site 
to HH1-site are calculated to be 0.927 (eV) and 1.001 (eV), respectively, whereas that from HH1-site to 
hH-site are more than 1.001 (eV). Furthermore, there is a large barrier energy of 1.829 (eV) from HH1-site 
to HH2-site, which is energetically unfavorable. But the energy barrier along HH2-site to HH1-site is 
about 0.284 (eV), that is related to the reduction of the density of graphyne layers in this area 
\cite{zhang2011high} \cite{xu2016promising}.

\section{CONCLUSION}

In this work, density functional calculations were employed to systematically investigate the pristine 
graphyne and Mg-decorated graphyne structures as a new anode material of magnesium-ion batteries, using 
Quantum Espresso package. The obtained results from calculated adsorption energy and adsorption height 
determined the most stable adsorption sites of magnesium on graphyne. Moreover, exploring on the DOS and 
band structure of the pristine graphyne and Mg-decorated graphyne structures, it was observed that the 
Mg adsorption makes graphyne metallic. we calculated a direct band gap of 0.358 (eV) for pristine 
graphyne, which is in good agreement with the results of 0.38 (eV) for Na-decorated graphyne batteries 
from other DFT calculations \cite{sarkar2015first}. Also, we determined the Mg-migration behaviors on 
graphyne monolayer on the way to the most stable adsorption sites for two main in-plane and out-plane 
diffusion paths, in addition, an out-plane diffusion pathway of Mg in bulk graphyne along the 
Hh-hH-HH1-HH2 direction, using Nudged Elastic Band method. The activation barrier of this process for 
in-plane diffusion path along h1-H1, H1-H2, and H2-h2 directions are 0.072 (eV), 0.835 (eV), and 
0.783 (eV), respectively. Besides, the energy barrier for Mg out-plane diffusion through H-site and 
h-site is about 0.32 (eV) and 15.68 (eV), respectively. Furthermore, for diffusion of Mg in bulk graphyne 
along the Hh-hH, hH-HH1, and HH1-HH2 paths, energy barriers are calculated to be 0.927 (eV), 1.001 (eV), 
and 1.829 (eV), respectively. In summary, investigation of details about Mg intercalation and diffusion 
into graphyne monolayer is highly desirable to shed light on this topic and to achieve the better 
understanding of battery technology.

\section{ACKNOWLEDGMENTS}
 The authors gratefully acknowledge the Sheikh Bahaei National High Performance Computing Center (SBNHPCC) for 
providing computing facilities and time. SBNHPCC is supported by scientific and technological department of presidential 
office and Isfahan University of Technology (IUT).


\bibliography{Gy}
\end{document}